\documentclass[lettersize,journal]{IEEEtran}
\usepackage{amsmath,amsfonts}
\usepackage{algorithm}
\usepackage{array}
\usepackage[caption=false,font=footnotesize,labelfont=rm,textfont=rm]{subfig}
\usepackage{textcomp}
\usepackage{stfloats}
\usepackage{url}
\usepackage{verbatim}
\usepackage{graphicx}
\usepackage{cite}
\usepackage{amssymb}
\usepackage{booktabs}
\usepackage[colorlinks,linkcolor=blue,anchorcolor=blue,citecolor=blue]{hyperref} 
\usepackage{xcolor}
\usepackage{orcidlink}
\usepackage[noend]{algpseudocode}

\begin{document}

\title{\fontsize{16}{19.2}\selectfont{Dual Orthogonal Projections for Multiuser Interference Cancellation in mmWave Beamforming With Uniform Planar Arrays}}

\author{Jiazhe Li$^{\orcidlink{0000-0002-1042-9981}}$,~\IEEEmembership{Member,~IEEE}, Heng Dong$^{\orcidlink{0000-0001-8961-0531}}$,~\IEEEmembership{Member,~IEEE}, Nicolò Decarli$^{\orcidlink{0000-0002-0359-8808}}$,~\IEEEmembership{Member,~IEEE},

Francesco Guidi$^{\orcidlink{0000-0002-1773-8541}}$,~\IEEEmembership{Member,~IEEE}, Anna Guerra$^{\orcidlink{0000-0001-5214-1444}}$,~\IEEEmembership{Member,~IEEE},

Alessandro Bazzi$^{\orcidlink{0000-0003-3500-1997}}$,~\IEEEmembership{Senior Member,~IEEE} and Zhuoming Li$^{\orcidlink{0000-0001-5172-6256}}$
        
\thanks{This work was supported by the China Scholarship Council under Grant No.202406120095.}

\thanks{\textit{(Corresponding author: Heng Dong.)}}

\thanks{Jiazhe Li, Heng Dong and Zhuoming Li are with the Department of Electronics and Information Engineering, Harbin Institute of Technology, Harbin 150001, China (e-mail: jiazhe@stu.hit.edu.cn; dongheng@hit.edu.cn; zhuoming@hit.edu.cn).}

\thanks{Nicolò Decarli and Francesco Guidi are with the National Research Council, Institute of Electronics, Computer and Telecommunication Engineering (CNR-IEIIT) and WiLab-CNIT, 40136 Bologna, Italy (e-mail: nicolo.decarli@cnr.it; francesco.guidi@cnr.it).}

\thanks{Anna Guerra and Alessandro Bazzi are with the DEI, University of Bologna and WiLab-CNIT, 40136 Bologna, Italy (e-mail: anna.guerra3@unibo.it; alessandro.bazzi@unibo.it).}
}




\maketitle

\begin{abstract}
This paper investigates multiuser interference (MUI) cancellation for millimeter-wave (mmWave) beamforming in communication systems. We propose a linear algorithm, termed iterative dual orthogonal projections (DOP), which alternates between two orthogonal projections: one to eliminate MUI and the other to refine combiners, ensuring empirical convergence in spectral efficiency. Simulation results show that, with each iteration, the spectral efficiency of each user converges rapidly, closely approaching the theoretical optimum determined by dirty paper coding (DPC), surpassing existing linear benchmarks while maintaining low computational complexity. Furthermore, the proposed DOP algorithm is extended to support both fully-digital and hybrid beamforming architectures.
\end{abstract}


\section{Introduction}
\label{secI}

\IEEEPARstart{A}{s} wireless communications evolve from sub-6 GHz to the mmWave band, antenna miniaturization facilitates the deployment of uniform planar arrays (UPAs), where beamforming techniques are widely adopted to enhance spectral efficiency \cite{zilli2021constrained}. However, multiuser beamforming suffers from significant multiuser interference (MUI). Although nonlinear algorithms achieve spectral efficiency close to the theoretical optimum defined by dirty paper coding (DPC) \cite{jindal2005sum}, their computational complexity is often prohibitive. This renders low-complexity linear algorithms a more practical alternative.

In this context, various linear algorithms have been proposed for multiuser beamforming. Specifically, the eigen zero-forcing (EZF) algorithm \cite{sun2010eigen} suppresses MUI with minimal computational overhead, whereas regularized block diagonalization (RBD) \cite{stankovic2008generalized} improves spectral efficiency by tolerating residual MUI. To further enhance performance, the WMMSE algorithm \cite{shi2011iteratively} achieves near-optimal spectral efficiency via the alternating optimization of beamformers and auxiliary variables, yet at the cost of prohibitive computational complexity. While these algorithms are primarily designed for fully-digital arrays, hybrid array architectures \cite{el2014spatially,yu2016alternating,gong2020rf} have emerged to reduce hardware costs, albeit at the expense of additional performance degradation \cite{khalid2019hybrid,li2025aree}. Consequently, developing low-complexity linear algorithms that closely approach the theoretical optimum remains a critical challenge for both fully-digital and hybrid beamforming systems.

To bridge this gap, we propose a linear multiuser beamforming algorithm, termed iterative dual orthogonal projections (DOP). It alternates between two orthogonal projections: one eliminates MUI, and the other refines combiners. DOP differs fundamentally from existing schemes, such as WMMSE, in its core design principle: The combiner obtained after MUI elimination may still contain ``redundant components'' that do not contribute to beamforming gain, which can be replaced with more effective counterparts through a refinement process. This iterative procedure demonstrates empirical convergence in spectral efficiency, outperforming existing linear benchmarks and closely approaching the DPC-based theoretical optimum. Furthermore, it can be extended to support both fully-digital and hybrid array architectures. Notably, both DOP and its hybrid extension remain suboptimal solutions, as the full potential of hybrid architectures merits further investigation.

\section{System Model}
\label{secII}

\begin{figure}[t]
  \centering
  \includegraphics[width=1\columnwidth]{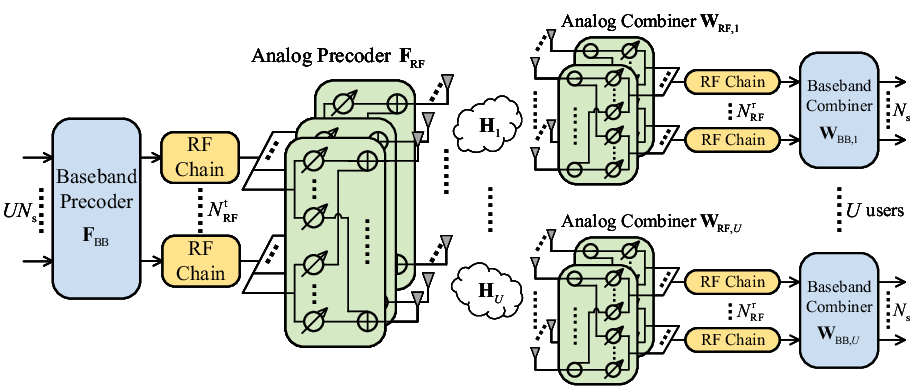}
  \caption{Downlink multiuser beamforming in mmWave UPA systems with fully-connected hybrid array architectures.}
  \label{fig1}
\end{figure}

As illustrated in Fig. \ref{fig1}, we consider a downlink multiuser mmWave UPA communication system, where the base station (BS) and mobile stations (MSs) adopt fully-connected hybrid array architectures \cite{el2014spatially}. The BS is equipped with ${{N}_{\text{t}}}$ antennas and $N_{\text{RF}}^{\text{t}}$ radio frequency (RF) chains to simultaneously serve $U$ MSs. Each MS is configured with ${{N}_{\text{r}}}$ antennas and $N_{\text{RF}}^{\text{r}}$ RF chains, supporting the transmission of ${{N}_{\text{s}}}$ data streams. This implies that the BS handles $U{{N}_{\text{s}}}$ data streams concurrently. Moreover, $U{{N}_{\text{s}}}\le N_{\text{RF}}^{\text{t}}\ll {{N}_{\text{t}}}$ and ${{N}_{\text{s}}}\le N_{\text{RF}}^{\text{r}}\ll {{N}_{\text{r}}}$ are satisfied to ensure the communication feasibility.

At the BS, the data streams are combined as $\mathbf{s}={{\left[ \mathbf{s}_{1}^{T},\ldots ,\mathbf{s}_{U}^{T} \right]}^{T}}\in {{\mathbb{C}}^{U{{N}_{\text{s}}}\times 1}}$ where ${{\mathbf{s}}_{u}}\in {{\mathbb{C}}^{{{N}_{\text{s}}}\times 1}}$ denotes the data symbol transmitted to the $u$-th MS, with $\mathbb{E}\left[ \mathbf{s}{{\mathbf{s}}^{H}} \right]=\left( {{P}}/{\left( U{{N}_{\text{s}}} \right)}\; \right){{\mathbf{I}}_{U{{N}_{\text{s}}}}}$, where ${P}$ represents the total transmit power. The baseband signal $\mathbf{s}$ is first processed by the baseband digital precoder ${{\mathbf{F}}_{\text{BB}}}=\left[ {{\mathbf{F}}_{\text{BB,1}}},\ldots ,{{\mathbf{F}}_{\text{BB,}U}} \right]\in {{\mathbb{C}}^{N_{\text{RF}}^{\text{t}}\times U{{N}_{\text{s}}}}}$, where ${{\mathbf{F}}_{\text{BB,}u}}$ is the digital precoder for the $u$-th MS. The signal is then up-converted via the RF analog precoder ${{\mathbf{F}}_{\text{RF}}}\in {{\mathbb{C}}^{{{N}_{\text{t}}}\times N_{\text{RF}}^{\text{t}}}}$ before being transmitted through the antenna array. Similarly, at the $u$-th MS, the received signal is first down-converted to baseband via the RF analog combiner ${{\mathbf{W}}_{\text{RF},u}}\in {{\mathbb{C}}^{{{N}_{\text{r}}}\times N_{\text{RF}}^{\text{r}}}}$, followed by the baseband digital combiner ${{\mathbf{W}}_{\text{BB},u}}\in {{\mathbb{C}}^{N_{\text{RF}}^{\text{r}}\times {{N}_{\text{s}}}}}$ to obtain the final received signal as
\begin{equation}
\begin{aligned}
  & {{\mathbf{y}}_{u}}=\mathbf{W}_{\text{BB},u}^{H}\mathbf{W}_{\text{RF},u}^{H}{{\mathbf{H}}_{u}}{{\mathbf{F}}_{\text{RF}}}{{\mathbf{F}}_{\text{BB}}}\mathbf{s}+\mathbf{W}_{\text{BB},u}^{H}\mathbf{W}_{\text{RF},u}^{H}{{\mathbf{n}}_{u}} \\ 
 & \quad \,=\mathbf{W}_{\text{BB},u}^{H}\mathbf{W}_{\text{RF},u}^{H}{{\mathbf{H}}_{u}}{{\mathbf{F}}_{\text{RF}}}{{\mathbf{F}}_{\text{BB,}u}}{{\mathbf{s}}_{u}}+\mathbf{W}_{\text{BB},u}^{H}\mathbf{W}_{\text{RF},u}^{H}{{\mathbf{n}}_{u}} \\ 
 & \quad \quad \quad \quad \quad \quad \quad \; \; +\underbrace{\sum\limits_{v\ne u}{\mathbf{W}_{\text{BB},u}^{H}\mathbf{W}_{\text{RF},u}^{H}{{\mathbf{H}}_{u}}{{\mathbf{F}}_{\text{RF}}}{{\mathbf{F}}_{\text{BB,}v}}{{\mathbf{s}}_{v}}}}_{\text{MUI}} \;, \\
\end{aligned}
\label{e1}
\end{equation}
where ${{\mathbf{H}}_{u}}\in {{\mathbb{C}}^{{{N}_{\text{r}}}\times {{N}_{\text{t}}}}}$ denotes the downlink mmWave channel matrix from the BS to the $u$-th MS, as detailed in \cite{li2025aree}. ${{\mathbf{n}}_{u}}\sim \mathcal{C}\mathcal{N}\left( \mathbf{0},\sigma _\mathrm{n}^{2}{{\mathbf{I}}_{{{N}_{\text{r}}}}} \right)$ is the additive white Gaussian noise (AWGN) for the $u$-th MS. In addition, ${{\mathbf{F}}_{\text{BB}}}$ is normalized to satisfy the BS power constraint as $\left\| {{\mathbf{F}}_{\text{RF}}}{{\mathbf{F}}_{\text{BB}}} \right\|_{F}^{2}=U{{N}_{\text{s}}}$. Both ${{\mathbf{F}}_{\text{RF}}}$ and ${{\mathbf{W}}_{\text{RF},u}}$ are subject to constant modulus constraints \cite{el2014spatially} such that $\left| {{\left[ {{\mathbf{F}}_{\text{RF}}} \right]}_{i,j}} \right|={1}/{\sqrt{{{N}_{\text{t}}}}}\;$ and $\left| {{\left[ {{\mathbf{W}}_{\text{RF},u}} \right]}_{i,j}} \right|={1}/{\sqrt{{{N}_{\text{r}}}}}\;$.

\section{DOP Algorithm for Multiuser Beamforming}
\label{secIII}

We first consider a fully-digital array architecture in this section, where the hybrid precoder ${{\mathbf{F}}_{\text{RF}}}{{\mathbf{F}}_{\text{BB,}u}}$ and combiner ${{\mathbf{W}}_{\text{RF},u}}{{\mathbf{W}}_{\text{BB},u}}$ are replaced with the fully-digital precoder ${{\mathbf{F}}_{u}}\in {{\mathbb{C}}^{{{N}_{\text{t}}}\times {{N}_{\text{s}}}}}$ and combiner ${{\mathbf{W}}_{u}}\in {{\mathbb{C}}^{{{N}_{\text{r}}}\times {{N}_{\text{s}}}}}$, respectively. Under this setup, the signal model (\ref{e1}) simplifies to
\begin{equation}
  {{\mathbf{y}}_{u}}=\mathbf{W}_{u}^{H}{{\mathbf{H}}_{u}}{{\mathbf{F}}_{u}}{{\mathbf{s}}_{u}}+\underbrace{\sum\limits_{v\ne u}{\mathbf{W}_{u}^{H}{{\mathbf{H}}_{u}}{{\mathbf{F}}_{v}}{{\mathbf{s}}_{v}}}}_{\text{MUI}}+\mathbf{W}_{u}^{H}{{\mathbf{n}}_{u}}.
\label{e2}
\end{equation}
Assuming Gaussian symbols are transmitted over the mmWave channel, the spectral efficiency of the $u$-th MS is given by \cite{stankovic2008generalized}:
\begin{equation}
  {{R}_{u}}\!=\!{{\log }_{2}}\left( \left| {{\mathbf{I}}_{{{N}_{\text{s}}}}}+\frac{{P}}{U{{N}_{\text{s}}}}\mathbf{R}_{u}^{-1}\mathbf{W}_{u}^{H}{{\mathbf{H}}_{u}}{{\mathbf{F}}_{u}}\mathbf{F}_{u}^{H}\mathbf{H}_{u}^{H}{{\mathbf{W}}_{u}} \right| \right),
\label{e3}
\end{equation}
where ${{\mathbf{R}}_{u}}\in {{\mathbb{C}}^{{{N}_{\text{s}}}\times {{N}_{\text{s}}}}}$ denotes the covariance matrix of the combined MUI and AWGN for the $u$-th MS, expressed as
\begin{equation}
  {{\mathbf{R}}_{u}}\!=\!\frac{{P}}{U\!{{N}_{\text{s}}}}\!\sum\limits_{v=1,v\ne u}^{U}{\!\!\!\!\!\!\left( \mathbf{W}_{u}^{H}{{\mathbf{H}}_{u}}{{\mathbf{F}}_{v}}\mathbf{F}_{v}^{H}\mathbf{H}_{u}^{H}{{\mathbf{W}}_{u}} \right)}+\sigma _{n}^{2}\mathbf{W}_{u}^{H}{{\mathbf{W}}_{u}}.
\label{e4}
\end{equation}
To maximize ${{R}_{\text{total}}}=\!\sum\nolimits_{u=1}^{U}{\!{{R}_{u}}}$, we design ${{\mathbf{F}}_{u}}$ and ${{\mathbf{W}}_{u}}$ using the proposed DOP algorithm, which iteratively alternates between two orthogonal projections as follows.

\subsection{Eliminating MUI via the First Orthogonal Projection}
Assume that the $u$-th user adopts a two-stage fully-digital combiner ${{\mathbf{W}}_{u}}={{\mathbf{W}}_{u,1}}{{\mathbf{W}}_{u,2}}$, where ${{\mathbf{W}}_{u,1}}=\mathbf{W}_{u}^{\text{initial}}\in {{\mathbb{C}}^{{{N}_{\text{r}}}\times {{N}_{\text{s}}}}}$ may be randomly initialized for the first iteration. The effective channel processed by ${{\mathbf{W}}_{u,1}}$ is given by
\begin{equation}
  {{\mathbf{\tilde{H}}}_{u}}=\mathbf{W}_{u,1}^{H}{{\mathbf{H}}_{u}}\in {{\mathbb{C}}^{{{N}_{\text{s}}}\times {{N}_{\text{t}}}}},\quad u=1,\ldots ,U.
\label{e5}
\end{equation}
Subsequently, excluding the $u$-th user, the remaining $U-1$ users form the following composite channel
\begin{equation}
  {{\left(\! \mathbf{\tilde{H}}_{u}^{\text{com}} \!\right)}^{\!\!H}}\!=\!\left[ \mathbf{\tilde{H}}_{1}^{H},\ldots ,\mathbf{\tilde{H}}_{u-1}^{H},\mathbf{\tilde{H}}_{u+1}^{H},\ldots ,\mathbf{\tilde{H}}_{U}^{H} \right]\!\in {{\mathbb{C}}^{{{N}_{\text{t}}}\times \left( U-1 \right){{N}_{\text{s}}}}}.
\label{e6}
\end{equation}
To eliminate MUI, we first calculate the column space of ${{\left(\! \mathbf{\tilde{H}}_{u}^{\text{com}} \!\right)}^{\!\!H}}$, followed by null space construction via orthogonal projection. Specifically, we define $\operatorname{span}\left( \cdot \right)$ as 
the span operator, and let
\begin{equation}
  \begin{aligned}
  & \operatorname{span}\left( {{\mathbf{F}}_{\text{I},u}} \right)=\operatorname{span}\left( {{\left( \mathbf{\tilde{H}}_{u}^{\text{com}} \right)}^{H}} \right) \\ 
  &\; \text{s}\text{.t}\text{.   }\mathbf{F}_{\text{I},u}^{H}{{\mathbf{F}}_{\text{I},u}}={{\mathbf{I}}_{{{L}_{u}}}}, \\ 
 \end{aligned}
\label{e7}
\end{equation}
where ${{\mathbf{F}}_{\text{I},u}}\in {{\mathbb{C}}^{{{N}_{\text{t}}}\times {{L}_{u}}}}$ is a semi-unitary matrix consisting of ${{L}_{u}}$ orthonormal basis vectors spanning the column space of ${{\left( \mathbf{\tilde{H}}_{u}^{\text{com}} \right)}^{H}}$. Consequently, the null space of ${{\left( \mathbf{\tilde{H}}_{u}^{\text{com}} \right)}^{H}}$ can be obtained by projecting ${{\mathbf{\tilde{H}}}_{u}}$ onto the orthogonal space of ${{\mathbf{F}}_{\text{I},u}}$
\begin{equation}
  {{\mathbf{\bar{H}}}_{u}}={{\mathbf{\tilde{H}}}_{u}}\left( {{\mathbf{I}}_{{{N}_{\text{t}}}}}-{{\mathbf{F}}_{\text{I},u}}\mathbf{F}_{\text{I},u}^{H} \right).
  \label{e8}
\end{equation}
Subsequently, applying SVD to ${{\mathbf{\bar{H}}}_{u}}$ as ${{\mathbf{\bar{H}}}_{u}}={{\mathbf{\bar{U}}}_{u}}{{\mathbf{\bar{\Sigma }}}_{u}}\mathbf{\bar{V}}_{u}^{H}$, we obtain the precoder and second-stage combiner that maximize spectral efficiency of ${{\mathbf{\bar{H}}}_{u}}$ as follows
\begin{equation}
  {{\mathbf{F}}_{u}}={{\left[ {{{\mathbf{\bar{V}}}}_{u}} \right]}_{:,1:{{N}_{\text{s}}}}}\in {{\mathbb{C}}^{{{N}_{\text{t}}}\times {{N}_{\text{s}}}}},
\label{e9}
\end{equation}
\begin{equation}
  {{\mathbf{W}}_{u,2}}={{\left[ {{{\mathbf{\bar{U}}}}_{u}} \right]}_{:,1:{{N}_{\text{s}}}}}\in {{\mathbb{C}}^{{{N}_{\text{s}}}\times {{N}_{\text{s}}}}},
\label{e10}
\end{equation}
where ${{\left[ \\ \cdot \\  \right]}_{i:j,:}}$(or ${{\left[ \\ \cdot \\  \right]}_{:,i:j}}$) is a matrix consisting of the elements from the $i$-th row (or column) to the $j$-th row (or column). The final combiner is thus constructed as ${{\mathbf{W}}_{u}}={{\mathbf{W}}_{u,1}}{{\mathbf{W}}_{u,2}}$.

The ${{\mathbf{F}}_{u}}$ and ${{\mathbf{W}}_{u}}$ derived above can completely eliminate MUI in (\ref{e2}), as demonstrated below. From (\ref{e8}), it is straightforward to verify that ${{\mathbf{\bar{H}}}_{u}}{{\mathbf{F}}_{\text{I},u}}=\mathbf{0}$. This leads to $\mathbf{\bar{V}}_{u}^{H}{{\mathbf{F}}_{\text{I},u}}=\mathbf{0}$, and by substituting (\ref{e9}), we obtain $\mathbf{F}_{u}^{H}{{\mathbf{F}}_{\text{I},u}}=\mathbf{0}$. Since ${{\mathbf{F}}_{\text{I},u}}$ spans the column space of ${{\left( \mathbf{\tilde{H}}_{u}^{\text{com}} \right)}^{H}}$ as illustrated in (\ref{e7}), it follows that $\mathbf{F}_{u}^{H}{{\left( \mathbf{\tilde{H}}_{u}^{\text{com}} \right)}^{H}}=\mathbf{0}$, which implies
\begin{equation}
  \mathbf{\tilde{H}}_{u}^{\text{com}}{{\mathbf{F}}_{u}}={{\left[ \mathbf{\tilde{H}}_{1}^{T},\ldots ,\mathbf{\tilde{H}}_{u-1}^{T},\mathbf{\tilde{H}}_{u+1}^{T},\ldots ,\mathbf{\tilde{H}}_{U}^{T} \right]}^{T}}{{\mathbf{F}}_{u}}=\mathbf{0}.
\label{e11}
\end{equation}
Therefore, we have ${{\mathbf{\tilde{H}}}_{v}}{{\mathbf{F}}_{u}}=\mathbf{W}_{v,1}^{H}{{\mathbf{H}}_{v}}{{\mathbf{F}}_{u}}=\mathbf{0}$ for any user $v\ne u$, which confirms that MUI is completely eliminated.

\subsection{Improving Spectral Efficiency via the Second Projection}
Although ${{\mathbf{F}}_{u}}$ and ${{\mathbf{W}}_{u}}$ obtained from the first projection can eliminate MUI, ${{\mathbf{W}}_{u}}$ may still contain ``redundant components" that are orthogonal to ${{\mathbf{H}}_{u}}{{\mathbf{F}}_{u}}$ and therefore do not contribute to the beamforming gain $\mathbf{W}_{u}^{H}{{\mathbf{H}}_{u}}{{\mathbf{F}}_{u}}$. \mbox{Consequently, ${{\mathbf{W}}_{u}}$ can be} further refined to enhance spectral efficiency. Specifically, let
\begin{equation}
  \begin{aligned}
  & \operatorname{span}\left( \mathbf{W}_{u}^{\text{eq}} \right)=\operatorname{span}\left( {{\mathbf{H}}_{u}}{{\mathbf{F}}_{u}} \right) \\ 
 &\; \text{s}\text{.t}\text{.   }{{\left( \mathbf{W}_{u}^{\text{eq}} \right)}^{H}}\mathbf{W}_{u}^{\text{eq}}={{\mathbf{I}}_{{{K}_{u}}}}, \\ 
\end{aligned}
\label{e12}
\end{equation}
where $\mathbf{W}_{u}^{\text{eq}}\in {{\mathbb{C}}^{{{N}_{\text{r}}}\times {{K}_{u}}}}$ is a semi-unitary matrix consisting of ${{K}_{u}}$ orthonormal basis vectors spanning the column space of ${{\mathbf{H}}_{u}}{{\mathbf{F}}_{u}}$. We refine ${{\mathbf{W}}_{u}}$ by projecting it onto $\mathbf{W}_{u}^{\text{eq}}$ as follows
\begin{equation}
  {{\mathbf{W}}_{u,1}}=\mathbf{W}_{u}^{\text{eq}}{{\left( \mathbf{W}_{u}^{\text{eq}} \right)}^{H}}{{\mathbf{W}}_{u}}\in {{\mathbb{C}}^{{{N}_{\text{r}}}\times {{N}_{\text{s}}}}},
\label{e13}
\end{equation}
where the redundant components (orthogonal to ${{\mathbf{H}}_{u}}{{\mathbf{F}}_{u}}$) are removed. The refined combiner ${{\mathbf{W}}_{u,1}}$ comprises two types of components that both contribute to the beamforming gain (i.e., non-orthogonal to ${{\mathbf{H}}_{u}}{{\mathbf{F}}_{u}}$): (i) the components from ${{\mathbf{W}}_{u}}$ which are non-orthogonal to $\mathbf{W}_{u}^{\text{eq}}$, and (ii) the newly introduced components from $\mathbf{W}_{u}^{\text{eq}}$. As a result, the redundant components in ${{\mathbf{W}}_{u}}$ are replaced by those in $\mathbf{W}_{u}^{\text{eq}}$, which contribute more effectively to the beamforming gain, as detailed below. 

According to (\ref{e12}), there exists an invertible matrix $\mathbf{A}$ such that ${{\mathbf{H}}_{u}}{{\mathbf{F}}_{u}}=\mathbf{W}_{u}^{\text{eq}}\mathbf{A}$. It then follows that
\begin{equation}
  \mathbf{W}_{u,1}^{H}{{\mathbf{H}}_{u}}{{\mathbf{F}}_{u}}\!=\!\mathbf{W}_{u}^{H}\mathbf{W}_{u}^{\text{eq}}{{\left( \mathbf{W}_{u}^{\text{eq}} \right)}^{\!H}}\mathbf{W}_{u}^{\text{eq}}\mathbf{A}\!=\!\mathbf{W}_{u}^{H}{{\mathbf{H}}_{u}}{{\mathbf{F}}_{u}}.
  \label{e14}
\end{equation}
Since ${{\left\| {{\mathbf{W}}_{u,1}} \right\|}_{F}}\le {{\left\| {{\mathbf{W}}_{u}} \right\|}_{F}}$ can be derived from (\ref{e12}) and (\ref{e13}), the following inequality is then established based on (\ref{e14}):
\begin{equation}
  {{\left\| {{\left( \frac{{{\mathbf{W}}_{u,1}}}{{{\left\| {{\mathbf{W}}_{u,1}} \right\|}_{F}}} \right)}^{H}}{{\mathbf{H}}_{u}}{{\mathbf{F}}_{u}} \right\|}_{F}}\ge {{\left\| {{\left( \frac{{{\mathbf{W}}_{u}}}{{{\left\| {{\mathbf{W}}_{u}} \right\|}_{F}}} \right)}^{H}}{{\mathbf{H}}_{u}}{{\mathbf{F}}_{u}} \right\|}_{F}},
\label{e15}
\end{equation}
which confirms that ${\mathbf{W}}_{u,1}$ provides a more effective contribution to the beamforming gain than ${\mathbf{W}}_{u}$. Consequently, by substituting (\ref{e13}) back into (\ref{e5}) and iteratively re-optimizing (\ref{e5})--(\ref{e13}), the spectral efficiency can be further enhanced, as will be theoretically justified in Sec.~\ref{secIV-B}.

In summary, the DOP algorithm employs a linear iterative procedure (\ref{e5})--(\ref{e13}) to optimize ${{\mathbf{F}}_{u}}$ and ${{\mathbf{W}}_{u}}$. In each iteration, ${{\mathbf{W}}_{u,1}}$ is used to derive ${{\mathbf{F}}_{u}}$ and ${{\mathbf{W}}_{u,2}}$, and subsequently ${{\mathbf{W}}_{u}}$ is projected to obtain an updated ${{\mathbf{W}}_{u,1}}$ for the next loop. Consequently, the spectral efficiency improves monotonically as ${{\mathbf{F}}_{u}}$ and ${{\mathbf{W}}_{u}}$ are iteratively updated. Upon convergence, the spectral efficiency can be further enhanced via water-filling method \cite{spencer2004zero,jindal2005sum} as ${{\mathbf{F}}_{\text{opt}}}=\left[ {{\mathbf{F}}_{1}},\ldots ,{{\mathbf{F}}_{U}} \right]{{\mathbf{\Lambda }}^{{1}/{2}\;}}$, where $\mathbf{\Lambda }\in {{\mathbb{C}}^{U{{N}_{\text{s}}}\times U{{N}_{\text{s}}}}}$ is the power allocation matrix as defined in \cite{jindal2005sum}. The complete DOP is summarized in \textbf{Algorithm 1}, with $\mathbf{W}_{u}^{\text{initial}}$ denoting the given initialization.

\begin{algorithm}[t]
  \caption{DOP Algorithm for Multiuser Beamforming}\label{algorithm2}
  \begin{algorithmic}[1]
      \Require{${{\mathbf{H}}_{u}}\in {{\mathbb{C}}^{{{N}_{\text{r}}}\times {{N}_{\text{t}}}}}$ for $u=1,\ldots ,U$ and ${{N}_{\text{s}}}$}
      \State
      \textbf{Initialization:} 
      \State
      \quad \ ${{\mathbf{W}}_{u,1}}=\mathbf{W}_{u}^{\text{initial}}\in {{\mathbb{C}}^{{{N}_{\text{r}}}\times {{N}_{\text{s}}}}}$ for $u=1,\ldots ,U$;
      \State
      \textbf{repeat:}
      \State
      \quad \ ${{\mathbf{\tilde{H}}}_{u}}=\mathbf{W}_{u,1}^{H}{{\mathbf{H}}_{u}}$ for $u=1,\ldots ,U$;
      \State
      \quad \ \textbf{for} $u=1:U$ \textbf{do}
      \State
      \quad \ \quad \ ${{\left( \!\mathbf{\tilde{H}}_{u}^{\text{com}} \!\right)}^{\!\!H}}=\left[ \mathbf{\tilde{H}}_{1}^{H},\ldots ,\mathbf{\tilde{H}}_{u-1}^{H},\mathbf{\tilde{H}}_{u+1}^{H},\ldots ,\mathbf{\tilde{H}}_{U}^{H} \right]$;
      \State
        \quad \ \quad \ $\operatorname{span}\!\left( {{\mathbf{F}}_{\text{I},u}} \right)\!=\!\operatorname{span}\!\left(\! {{\left( \mathbf{\tilde{H}}_{u}^{\text{com}} \right)}^{\!\!H}} \!\right)\text{, s}\text{.t}\text{. }\mathbf{F}_{\text{I},u}^{H}{{\mathbf{F}}_{\text{I},u}}\!=\!{{\mathbf{I}}_{{{L}_{u}}}}$;
      \State
        \quad \ \quad \ ${{\mathbf{\bar{H}}}_{u}}={{\mathbf{\tilde{H}}}_{u}}\left( {{\mathbf{I}}_{{{N}_{\text{t}}}}}-{{\mathbf{F}}_{\text{I},u}}\mathbf{F}_{\text{I},u}^{H} \right)$;
      \State
        \quad \ \quad \ ${{\mathbf{\bar{H}}}_{u}}={{\mathbf{\bar{U}}}_{u}}{{\mathbf{\bar{\Sigma }}}_{u}}\mathbf{\bar{V}}_{u}^{H}$;
      \State
        \quad \ \quad \ ${{\mathbf{F}}_{u}}={{\left[ {{{\mathbf{\bar{V}}}}_{u}} \right]}_{:,1:{{N}_{\text{s}}}}}$, ${{\mathbf{W}}_{u,2}}={{\left[ {{{\mathbf{\bar{U}}}}_{u}} \right]}_{:,1:{{N}_{\text{s}}}}}$;
      \State
        \quad \ \quad \ ${{\mathbf{W}}_{u}}={{\mathbf{W}}_{u,1}}{{\mathbf{W}}_{u,2}}$;
      \State
        \quad \ \quad \ $\operatorname{span}\!\left( \mathbf{W}_{u}^{\text{eq}} \right)\!=\!\operatorname{span}\!\left( {{\mathbf{H}}_{u}}{{\mathbf{F}}_{\!u}} \right)\text{, s}\text{.t}\text{. }\!{{\left( \mathbf{W}_{u}^{\text{eq}} \right)}^{\!\!H}}\!\mathbf{W}_{u}^{\text{eq}}\!=\!{{\mathbf{I}}_{{{\!K}_{\!u}}}}$;      
      \State
        \quad \ \quad \ ${{\mathbf{W}}_{u,1}}=\mathbf{W}_{u}^{\text{eq}}{{\left( \mathbf{W}_{u}^{\text{eq}} \right)}^{H}}{{\mathbf{W}}_{u}}$;
      \State
        \quad \ \textbf{end for}
      \State
      \textbf{until} a stopping criterion triggers
      \State
      Calculate $\mathbf{\Lambda }$ using water-filling power allocation \cite{jindal2005sum};
      \State
      ${{\mathbf{F}}_{\text{opt}}}=\left[ {{\mathbf{F}}_{1}},\ldots ,{{\mathbf{F}}_{U}} \right]{{\mathbf{\Lambda }}^{{1}/{2}\;}}\in {{\mathbb{C}}^{{{N}_{\text{t}}}\times U{{N}_{\text{s}}}}}$;
      \State
      $\mathbf{W}_{u}^{\text{opt}}={{\mathbf{W}}_{u}}\in {{\mathbb{C}}^{{{N}_{\text{r}}}\times {{N}_{\text{s}}}}}$ for $u=1,\ldots ,U$;
      \Ensure
      ${{\mathbf{F}}_{\text{opt}}}$, $\mathbf{W}_{u}^{\text{opt}}$ for $u=1,\ldots ,U$
  \end{algorithmic}
\end{algorithm}

\subsection{Extending DOP Algorithm to Hybrid Beamforming}
\label{secIII-C}

Next, we consider a hybrid array architecture, which transforms (\ref{e2}) into (\ref{e1}). The hybrid beamforming problem can be equivalently reformulated as a matrix approximation problem to approximate the fully-digital solutions. Specifically, with the fully-digital solutions ${{\mathbf{F}}_{\text{opt}}}\in {{\mathbb{C}}^{{{N}_{\text{t}}}\times U{{N}_{\text{s}}}}}$ and $\mathbf{W}_{u}^{\text{opt}}\in {{\mathbb{C}}^{{{N}_{\text{r}}}\times {{N}_{\text{s}}}}}$ ($u=1,\ldots ,U$) obtained via \textbf{Algorithm~1}, the multiuser hybrid precoder design of ${{\mathbf{F}}_{\text{RF}}}\in {{\mathbb{C}}^{{{N}_{\text{t}}}\times N_{\text{RF}}^{\text{t}}}}$ and ${{\mathbf{F}}_{\text{BB}}}\in {{\mathbb{C}}^{N_{\text{RF}}^{\text{t}}\times U{{N}_{\text{s}}}}}$ can be formulated as
\begin{equation}
  \begin{aligned}
  & \underset{{{\mathbf{F}}_{\text{RF}}},{{\mathbf{F}}_{\text{BB}}}}{\mathop{\operatorname{minimize}}}\quad \left\| {{\mathbf{F}}_{\text{opt}}}-{{\mathbf{F}}_{\text{RF}}}{{\mathbf{F}}_{\text{BB}}} \right\|_{F}^{2} \\ 
 & \operatorname{subject \ to}\quad \left| {{\left[ {{\mathbf{F}}_{\text{RF}}} \right]}_{i,j}} \right|={1}/{\sqrt{{{N}_{\text{t}}}},\quad \forall i,j}\; \\ 
 & \quad \quad \quad \, \; \ \ \ \ \ \ \left\| {{\mathbf{F}}_{\text{RF}}}{{\mathbf{F}}_{\text{BB}}} \right\|_{F}^{2}=U{{N}_{\text{s}}} \ , \\ 
\end{aligned}
\label{e16}
\end{equation}
which has the same mathematical structure as the conventional single-user hybrid beamforming formulation. As a result, (\ref{e16}) can be effectively solved using established single-user linear algorithms such as AREE \cite{li2025aree} or PE-AltMin \cite{yu2016alternating} to achieve near-optimal solutions. Similarly, the hybrid combiner design of ${{\mathbf{W}}_{\text{RF},u}}\in {{\mathbb{C}}^{{{N}_{\text{r}}}\times N_{\text{RF}}^{\text{r}}}}$ and ${{\mathbf{W}}_{\text{BB},u}}\in {{\mathbb{C}}^{N_{\text{RF}}^{\text{r}}\times {{N}_{\text{s}}}}}$ for each user can be equivalently formulated as a matrix approximation problem to approximate $\mathbf{W}_{u}^{\text{opt}}$, which is identical in form to (\ref{e16}).

It is important to note that ${{\mathbf{F}}_{\text{RF}}}$, ${{\mathbf{F}}_{\text{BB}}}=\left[ {{\mathbf{F}}_{\text{BB,1}}},\ldots ,{{\mathbf{F}}_{\text{BB},U}} \right]$, ${{\mathbf{W}}_{\text{RF},u}}$ and ${{\mathbf{W}}_{\text{BB},u}}$ obtained from (\ref{e16}) inevitably introduce approximation errors, which in turn lead to MUI. Therefore, it is necessary to refine ${{\mathbf{F}}_{\text{BB,}u}}$ and ${{\mathbf{W}}_{\text{BB},u}}$ as follows. First, replace ${{\mathbf{\tilde{H}}}_{u}}$ in (\ref{e5}) with ${{\mathbf{\tilde{H}}}_{u}}=\mathbf{W}_{\text{BB},u}^{H}\mathbf{W}_{\text{RF},u}^{H}{{\mathbf{H}}_{u}}{{\mathbf{F}}_{\text{RF}}}$, and then execute procedures (\ref{e5})--(\ref{e10}). Next, update beamformers as ${{\mathbf{F}}_{\text{BB,}u}}={{\mathbf{F}}_{u}}$ and ${{\mathbf{W}}_{\text{BB},u}}={{\mathbf{W}}_{\text{BB},u}}{{\mathbf{W}}_{u,2}}$, where ${{\mathbf{F}}_{u}}$ and ${{\mathbf{W}}_{u,2}}$ are obtained from (\ref{e9}) and (\ref{e10}), respectively. Finally, perform the normalization ${{\mathbf{F}}_{\text{BB}}}=\left( {\sqrt{U{{N}_{\text{s}}}}}/{{{\left\| {{\mathbf{F}}_{\text{RF}}}{{\mathbf{F}}_{\text{BB}}} \right\|}_{F}}}\; \right){{\mathbf{F}}_{\text{BB}}}$ to satisfy the power constraint in (\ref{e16}). As a result, the MUI is further suppressed, following the same principle as DOP.

\section{Performance Analysis of DOP Algorithm}
\label{secIV}

\subsection{Complexity Analysis}
The computational complexity of the DOP algorithm is primarily dominated by steps (\ref{e5}), (\ref{e7}) and (\ref{e12}), which entail complexities of $\mathcal{O}\left( {{N}_{\text{t}}}{{N}_{\text{r}}}{{N}_{\text{s}}} \right)$, $\mathcal{O}\left( {{N}_{\text{t}}}N_{\text{s}}^{2}{{U}^{2}} \right)$ and $\mathcal{O}\left( {{N}_{\text{t}}}{{N}_{\text{r}}}{{N}_{\text{s}}} \right)$, respectively. Given that these procedures are executed for all $U$ users across ${{I}_{\text{iter}}}$ iterations, the total complexity of DOP is $\mathcal{O}\left( {{I}_{\text{iter}}}\left( {{N}_{\text{t}}}{{N}_{\text{r}}}{{N}_{\text{s}}}U+{{N}_{\text{t}}}N_{\text{s}}^{2}{{U}^{3}} \right) \right)$.

Regarding baseline schemes, the complexity of WMMSE \cite{shi2011iteratively} is $\mathcal{O}\left( {{I}_{\text{iter}}}\left( N_{\text{t}}^{3}+N_{\text{t}}^{2}{{N}_{\text{s}}}U \right) \right)$, which substantially exceeds that of DOP as it scales cubically with $N_\text{t}$. Meanwhile, RBD \cite{stankovic2008generalized} incurs a complexity of $\mathcal{O}\left( N_{\text{t}}^{2}{{N}_{\text{r}}}{{U}^{2}} \right)$. Notably, EZF \cite{sun2010eigen} exhibits the lowest complexity of $\mathcal{O}\left( {{N}_{\text{t}}}{{N}_{\text{r}}}{{N}_{\text{s}}}U+{{N}_{\text{t}}}N_{\text{s}}^{2}{{U}^{2}} \right)$, making it an ideal candidate to provide the initial value $\mathbf{W}_{u}^{\text{initial}}$ for \textbf{Algorithm 1}.

\subsection{Convergence Analysis}
\label{secIV-B}

Although Sec.~\ref{secV} shows empirical convergence for DOP, a rigorous theoretical proof is challenging. However, since ${{\left( \mathbf{F}_{\text{I},u}^{\left( t \right)} \right)}^{H}}\mathbf{F}_{u}^{\left( t \right)}=\mathbf{0}$ at $t$-th iteration, DOP can be proved to converge under the following approximation\footnote{This approximation is physically justified: it implies that the precoder designed to eliminate MUI at $t$-th iteration remains approximately effective at $(t+1)$-th iteration.}:
\begin{equation}
{{\left( \mathbf{F}_{\text{I},u}^{\left( t+1 \right)} \right)}^{H}}\mathbf{F}_{u}^{\left( t \right)}\approx \mathbf{0}.
\label{e17}
\end{equation} 
The proof is provided as follows.

\textit{Proof}: According to (\ref{e2}), the signal covariance matrix of the $u$-th user after the $t$-th iteration can be expressed as
\begin{equation}
  \begin{aligned}
  & \mathbf{S}_{u}^{\left( t \right)}={{\left( \mathbf{W}_{u}^{\left( t \right)} \right)}^{H}}{{\mathbf{H}}_{u}}\mathbf{F}_{u}^{\left( t \right)}{{\left( \mathbf{F}_{u}^{\left( t \right)} \right)}^{H}}\mathbf{H}_{u}^{H}\mathbf{W}_{u}^{\left( t \right)} \\ 
 & \quad \; \; \; \overset{(\text{a})}{\mathop{=}}\,{{\left( \mathbf{W}_{u,1}^{\left( t+1 \right)} \right)}^{H}}{{\mathbf{H}}_{u}}\mathbf{F}_{u}^{\left( t \right)}{{\left( \mathbf{F}_{u}^{\left( t \right)} \right)}^{H}}\mathbf{H}_{u}^{H}\mathbf{W}_{u,1}^{\left( t+1 \right)} \\ 
& \quad \; \; \; \overset{\left( \text{b} \right)}{\mathop{=}}\,\mathbf{\tilde{H}}_{u}^{\left( t+1 \right)}\mathbf{F}_{u}^{\left( t \right)}{{\left( \mathbf{F}_{u}^{\left( t \right)} \right)}^{H}}{{\left( \mathbf{\tilde{H}}_{u}^{\left( t+1 \right)} \right)}^{H}} \\
& \quad \; \; \; \overset{\left( \text{c} \right)}{\mathop{\approx }}\,\mathbf{\bar{H}}_{u}^{\left( t+1 \right)}\mathbf{F}_{u}^{\left( t \right)}{{\left( \mathbf{F}_{u}^{\left( t \right)} \right)}^{H}}{{\left( \mathbf{\bar{H}}_{u}^{\left( t+1 \right)} \right)}^{H}}, \\ 
\end{aligned}
\label{e18}
\end{equation}
where (a) and (b) follow from (\ref{e14}) and (\ref{e5}), respectively, and (c) is derived from (\ref{e8}) and approximation (\ref{e17}) as follows:
\begin{equation}
\mathbf{\bar{H}}_{u}^{\!\left( t+1\! \right)}\mathbf{F}_{u}^{\left( t \right)}\!=\mathbf{\tilde{H}}_{u}^{\!\left( t+1 \right)}\!\!\left(\! {{\mathbf{I}}_{{{\!N}_{\text{\!t}}}}}\!-\!\mathbf{F}_{\text{I},u}^{\!\left( t+1 \right)}\!{{\left(\! \mathbf{F}_{\text{I},u}^{\!\left( t+1 \right)} \!\right)}^{\!\!H}} \right)\!\mathbf{F}_{u}^{\left( t \right)}\!\approx \mathbf{\tilde{H}}_{u}^{\!\left( t+1 \!\right)}\mathbf{F}_{u}^{\left( t \right)}.
\label{e19}
\end{equation}
Similarly, $\mathbf{S}_{u}^{\left( t+1 \right)}$ after the $(t\!+\!1)$-th iteration is expressed as
\begin{small}
\begin{equation}
  \begin{aligned}
  & \mathbf{S}_{u}^{\left( t+\!1 \right)}\!=\!{{\left(\! \mathbf{W}_{u,2}^{\left( t+1 \right)} \!\right)}^{\!\!H}}\mathbf{\tilde{H}}_{u}^{\left( t+1 \right)}\mathbf{F}_{u}^{\left( t+1 \right)}{{\left(\! \mathbf{F}_{u}^{\left( t+1 \right)} \!\right)}^{\!\!H}}{{\left(\! \mathbf{\tilde{H}}_{u}^{\left( t+1 \right)} \!\right)}^{\!\!H}}\mathbf{W}_{u,2}^{\left( t+1 \right)} \\ 
 & \quad \ \ \ \,={{\left(\! \mathbf{W}_{u,2}^{\left( t+1 \right)} \!\right)}^{\!\!H}}\mathbf{\bar{H}}_{u}^{\left( t+1 \right)}\mathbf{F}_{u}^{\left( t+1 \right)}{{\left(\! \mathbf{F}_{u}^{\left( t+1 \right)} \!\right)}^{\!\!H}}{{\left(\! \mathbf{\bar{H}}_{u}^{\left( t+1 \right)} \!\right)}^{\!\!H}}\mathbf{W}_{u,2}^{\left( t+1 \right)}\\
 & \quad \ \; \; \; \succeq \mathbf{S}_{u}^{\left( t \right)}, \\ 
\end{aligned}
\label{e20}
\end{equation}
\end{small}where $\mathbf{S}_{u}^{\left( t+1 \right)}\succeq\mathbf{S}_{u}^{\left( t \right)}$ since $\mathbf{W}_{u,2}^{\left( t+1 \right)}$ and $\mathbf{F}_{u}^{\left( t+1 \right)}$ (derived from (\ref{e9}) and (\ref{e10})) are the left and right singular matrices corresponding to the largest ${{N}_{\text{s}}}$ singular values of $\mathbf{\bar{H}}_{u}^{\left( t+1 \right)}$. Consequently, $\operatorname{Tr}\left( \mathbf{S}_{u}^{\left( t+1 \right)} \right)\ge \operatorname{Tr}\left( \mathbf{S}_{u}^{\left( t \right)} \right)$ according to (\ref{e20}), which confirms that signal power of the $u$-th user increases monotonically under the approximation (\ref{e17}).

After the $t$-th iteration, MUI is fully eliminated by (\ref{e11}), leaving the $u$-th user subject only to AWGN. Therefore, (\ref{e4}) simplifies to $\mathbf{R}_{u}^{\left( t \right)}=\sigma _{\text{n}}^{2}{{\left( \mathbf{W}_{u}^{\left( t \right)} \right)}^{H}}\mathbf{W}_{u}^{\left( t \right)}$, which is then updated after the $(t+1)$-th iteration as
\begin{equation}
  \begin{aligned}
  & \mathbf{R}_{u}^{\left( t+1 \right)}=\sigma _{\text{n}}^{2}{{\left( \mathbf{W}_{u,2}^{\left( t+1 \right)} \right)}^{H}}{{\left( \mathbf{W}_{u,1}^{\left( t+1 \right)} \right)}^{H}}\mathbf{W}_{u,1}^{\left( t+1 \right)}\mathbf{W}_{u,2}^{\left( t+1 \right)} \\ 
 & \quad \ \ \ \; \; \; \preceq {{\left( \mathbf{W}_{u,2}^{\left( t+1 \right)} \right)}^{H}}\mathbf{R}_{u}^{\left( t \right)}\mathbf{W}_{u,2}^{\left( t+1 \right)}, \\ 
\end{aligned}
\label{e21}
\end{equation}
where the inequality is justified below. From (\ref{e13}), we have
\begin{equation}
\begin{aligned}
  & {{\left(\! \mathbf{W}_{u,1}^{\left( t+1 \right)} \!\right)}^{\!\!H}}\mathbf{W}_{u,1}^{\left( t+1 \right)}={{\left(\! \mathbf{W}_{u}^{\left( t \right)} \!\right)}^{\!\!H}}\mathbf{W}_{u}^{\text{eq}}{{\left( \mathbf{W}_{u}^{\text{eq}} \right)}^{H}}\mathbf{W}_{u}^{\left( t \right)} \\ 
 & \quad \quad \quad \quad \quad \quad \quad \ \ \preceq {{\left(\! \mathbf{W}_{u}^{\left( t \right)} \!\right)}^{\!\!H}}\mathbf{W}_{u}^{\left( t \right)}=\left( {1}/{\sigma _{n}^{2}}\; \right)\mathbf{R}_{u}^{\left( t \right)} \;, \\ 
\end{aligned}
\label{e22}
\end{equation}
where $\mathbf{W}_{u}^{\text{eq}}{{\left( \mathbf{W}_{u}^{\text{eq}} \right)}^{\!H}}\preceq {{\mathbf{I}}_{{{N}_{\text{r}}}}}$ (from (\ref{e12})) ensures (\ref{e22}) holds, thereby validating (\ref{e21}). Consequently, {\small$\operatorname{Tr}\left( \mathbf{R}_{u}^{\left( t+1 \right)} \right)\le \operatorname{Tr}\left( {{\left(\! \mathbf{W}_{u,2}^{\left( t+1 \right)} \!\right)}^{\!H}}\mathbf{R}_{u}^{\left( t \right)}\mathbf{W}_{u,2}^{\left( t+1 \right)} \right)=\operatorname{Tr}\left( \mathbf{R}_{u}^{\left( t \right)} \right)$}, with equality due to unitary property of $\mathbf{W}_{u,2}^{\left( t+1 \right)}$. Therefore, the equivalent noise power after receive combining for the $u$-th user decreases.

The spectral efficiency of the $u$-th user (defined in (\ref{e3})) is updated after the $t$-th iteration as
\begin{equation}
  \begin{small} 
  \begin{aligned}
  & R_{u}^{\left( t \right)}={{\log }_{2}}\left( \left| {{\mathbf{I}}_{{{N}_{\text{s}}}}}+\frac{{P}}{U{{N}_{\text{s}}}}{{\left( \mathbf{R}_{u}^{\left( t \right)} \right)}^{-1}}\mathbf{S}_{u}^{\left( t \right)} \right| \right) \\ 
 & \quad \ \ ={{\log }_{2}}\left( \left| {{\mathbf{I}}_{{{N}_{\text{s}}}}}+\frac{{P}}{U{{N}_{\text{s}}}}{{\left( \mathbf{W}_{u,2}^{\left( t+1 \right)} \right)}^{H}}{{\left( \mathbf{R}_{u}^{\left( t \right)} \right)}^{-1}}\mathbf{W}_{u,2}^{\left( t+1 \right)} \right. \right. \\ 
 & \left. \left. \quad \quad \quad \quad \quad \quad \quad \quad \quad \quad \quad \ \ \ \; \times {{\left( \mathbf{W}_{u,2}^{\left( t+1 \right)} \right)}^{H}}\mathbf{S}_{u}^{\left( t \right)}\mathbf{W}_{u,2}^{\left( t+1 \right)} \right| \right) \\ 
 & \quad \ \ \le {{\log }_{2}}\left( \left| {{\mathbf{I}}_{{{N}_{\text{s}}}}}+\frac{{P}}{U{{N}_{\text{s}}}}{{\left( \mathbf{R}_{u}^{\left( t+1 \right)} \right)}^{-1}}\mathbf{S}_{u}^{\left( t+1 \right)} \right| \right)=R_{u}^{\left( t+1 \right)}, \\ 
\end{aligned}
\label{e23}
\end{small}
\end{equation}
where the inequality holds since {\small${{\left( \mathbf{W}_{u,2}^{\left( t+1 \right)} \right)}^{H}}{{\left( \mathbf{R}_{u}^{\left( t \right)} \right)}^{-1}}\mathbf{W}_{u,2}^{\left( t+1 \right)}\preceq{{\left( \mathbf{R}_{u}^{\left( t+1 \right)} \right)}^{-1}}$} from (\ref{e21}), and ${{\left( \mathbf{W}_{u,2}^{\left( t+1 \right)} \right)}^{H}}\mathbf{S}_{u}^{\left( t \right)}\mathbf{W}_{u,2}^{\left( t+1 \right)}\preceq \mathbf{S}_{u}^{\left( t+1 \right)}$ from (\ref{e18}) and (\ref{e20}). As a result, the spectral efficiency of the $u$-th user increases monotonically, and the sum spectral efficiency of all users likewise increases. Since spectral efficiency is upper-bounded by the DPC-based theoretical optimum \cite{jindal2005sum}, the DOP algorithm is guaranteed to converge under the approximation (\ref{e17}). \hfill $\blacksquare$

\subsection{Performance under Imperfect CSI}
To evaluate the robustness of algorithms under imperfect channel state information (CSI), we adopt the following stochastic/Gaussian CSI uncertainty model \cite{rajput2022robust}:
\begin{equation}
  \mathbf{H}_u=\mathbf{\hat{H}}_u+\underbrace{\mathbf{R}_{\mathrm{r}}^{1/2\ }\mathbf{GR}_{\mathrm{t}}^{T/2\ }}_{\Delta \mathbf{H}}\ ,
\label{e24}
\end{equation}
where $\mathbf{H}_u$ is the ideal channel, $\mathbf{\hat{H}}_u$ represents the available channel estimate, and $\Delta \mathbf{H}$ denotes the estimation error. $\mathbf{G}\in {{\mathbb{C}}^{{{N}_{\mathrm{r}}}\times {{N}_{\mathrm{t}}}}}$ contains independent and identically distributed (i.i.d.) elements following $\mathcal{C}\mathcal{N}\left( 0,1 \right)$. The receive and transmit correlation matrices ${{\mathbf{R}}_{\mathrm{r}}}\in {{\mathbb{C}}^{{{N}_{\mathrm{r}}}\times {{N}_{\mathrm{r}}}}}$ and ${{\mathbf{R}}_{\mathrm{t}}}\in {{\mathbb{C}}^{{{N}_{\mathrm{t}}}\times {{N}_{\mathrm{t}}}}}$ are modeled as ${{\left[ {{\mathbf{R}}_{\mathrm{r}}} \right]}_{i,j}}=\sigma _{\mathrm{H}}^{2}\beta _{\mathrm{r}}^{\left| i-j \right|}$ and ${{\left[ {{\mathbf{R}}_{\mathrm{t}}} \right]}_{i,j}}=\beta _{\mathrm{t}}^{\left| i-j \right|}$, where ${{\beta }_{\mathrm{r}}}$ and ${{\beta }_{\mathrm{t}}}$ represent the receive and transmit correlation coefficients, and $\sigma _{\mathrm{H}}^{2}$ denotes the CSI uncertainty variance.

\begin{figure*}[t]
  \centering
  \subfloat[Sum spectral efficiency versus SNR]
  {
    \label{fig2a}\includegraphics[width=0.3\textwidth]{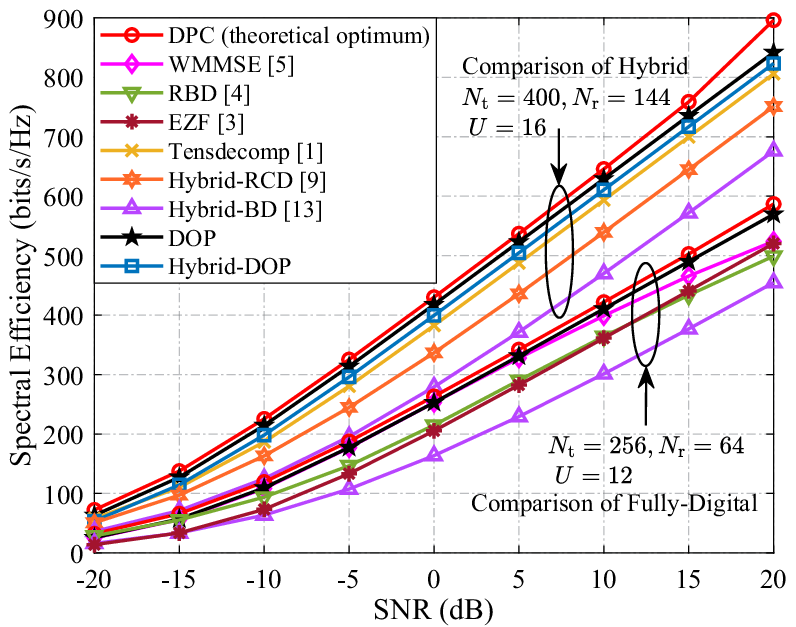}
  }
  \hfill
  \subfloat[Sum spectral efficiency versus user number]
  {
    \label{fig2b}\includegraphics[width=0.3\textwidth]{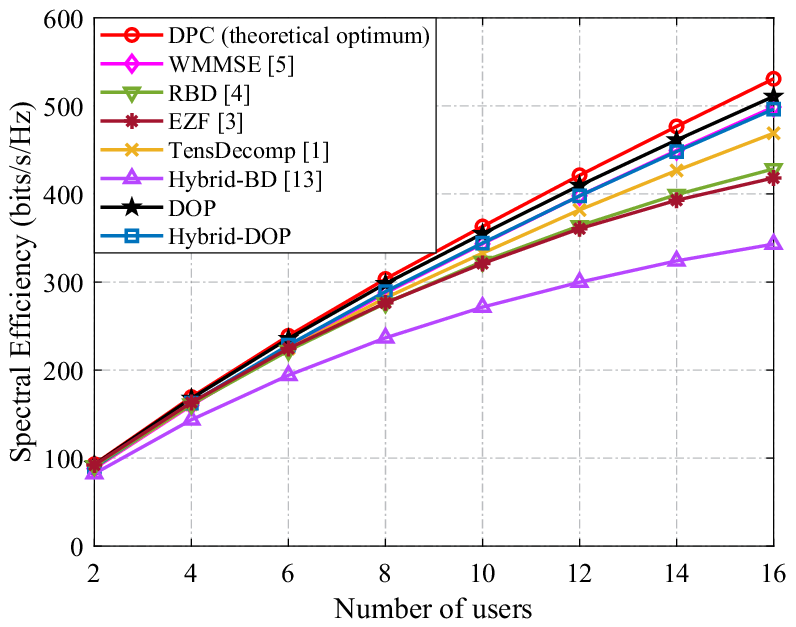}
  }
  \hfill
  \subfloat[Sum spectral efficiency under imperfect CSI]
  {
    \label{fig2c}\includegraphics[width=0.3\textwidth]{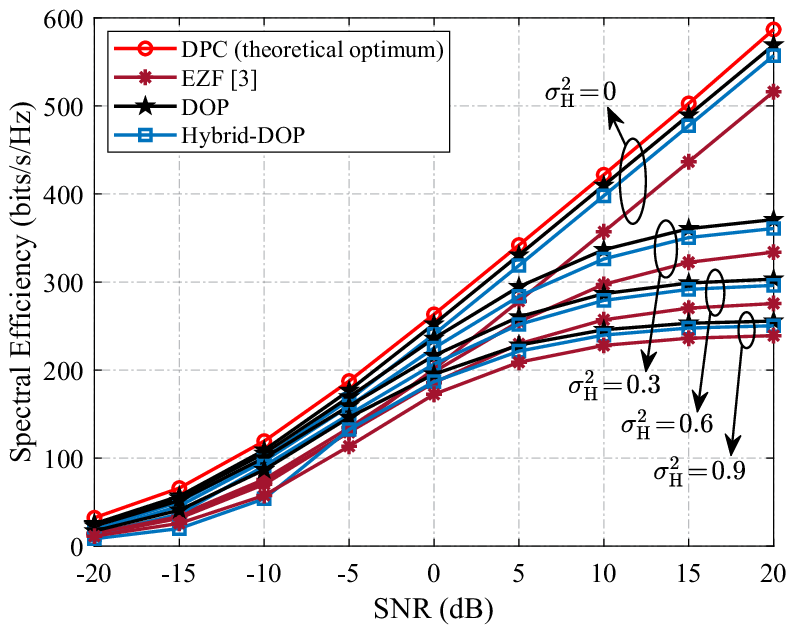}
  }
  \caption{Sum spectral efficiency versus different parameters for multiuser beamforming algorithms, where ${{N}_{\text{t}}}=256$, ${{N}_{\text{r}}}=64$, and ${{N}_{\text{s}}}=4$ for both fully-digital and hybrid architectures, while $N_{\text{RF}}^{\text{t}}=U{{N}_{\text{s}}}$ and $N_{\text{RF}}^{\text{r}}={{N}_{\text{s}}}$ apply to hybrid algorithms only. Fig. \ref{fig2}(b) evaluates performance at $\text{SNR}=\text{10 dB}$.}
\label{fig2}  
\end{figure*}

\section{Simulation Results}
\label{secV}

In this section, simulation results are provided to validate the performance. We include both existing linear fully-digital algorithms (EZF \cite{sun2010eigen}, RBD \cite{stankovic2008generalized} and WMMSE \cite{shi2011iteratively}) and hybrid algorithms (Hybrid-BD \cite{ni2015hybrid}, Hybrid-RCD \cite{khalid2019hybrid} and TensDecomp \cite{zilli2021constrained}), along with the theoretical optimum defined by DPC \cite{jindal2005sum} for comparison. All algorithm outputs are further optimized using water-filling \cite{spencer2004zero} to ensure a fair comparison. The DOP algorithm is executed for 20 iterations, with $\mathbf{W}_{u}^{\text{initial}}$ set to the output of the EZF algorithm. The SNR is defined as ${P}/{\sigma _{\text{n}}^{2}}\;$, and the channel parameters for $\mathbf{H}_u$ are adopted from \cite{li2025aree}. For imperfect CSI, we assume ${{\beta }_{\mathrm{r}}}={{\beta }_{\mathrm{t}}}=0.6$ as in \cite{rajput2022robust}. All simulation results are averaged over 1000 channel realizations.

\begin{figure}[t]
  \centering
  \subfloat[Impact of MUI refinement]
  {
    \label{fig3a}\includegraphics[width=0.23\textwidth]{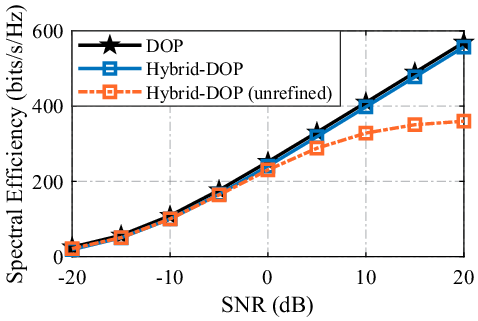}
  }
  \hfill
  \subfloat[Convergence of the $u$-th user]
  {
    \label{fig3b}\includegraphics[width=0.23\textwidth]{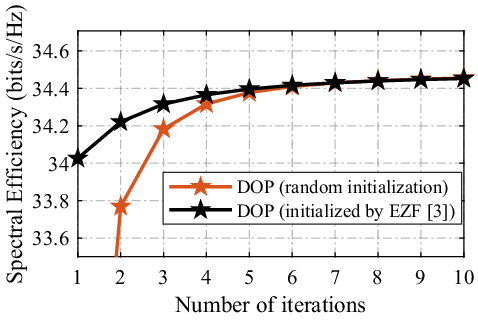}
  }
  \hfill
  \caption{Performance analysis regarding MUI refinement and convergence, where ${{N}_{\text{t}}}=256$, ${{N}_{\text{r}}}=64$, ${{N}_{\text{s}}}=4$, $U=12$, $N_{\text{RF}}^{\text{t}}=U{{N}_{\text{s}}}$ and $N_{\text{RF}}^{\text{r}}={{N}_{\text{s}}}$. Fig.~\ref{fig3}(b) evaluates performance at $\text{SNR}=\text{10 dB}$.}
  \label{fig3}
\end{figure}

Fig. \ref{fig2}(a) illustrates the sum spectral efficiency versus SNR, and Fig. \ref{fig2}(b) shows the performance versus the number of users. For fully-digital beamforming, the proposed DOP outperforms existing benchmarks and closely approaches the DPC-based theoretical optimum. However, DOP remains a suboptimal solution due to the inherent constraints of linear processing. In hybrid beamforming case, the proposed Hybrid-DOP first employs AREE \cite{li2025aree} to approximate the fully-digital solution of DOP, followed by refinements in Sec.~\ref{secIII-C} to further suppress MUI. It effectively approximates DOP and surpasses other hybrid schemes. As shown in Fig. \ref{fig3}(a), these MUI refinements for Hybrid-DOP are essential, as the unrefined version suffers significant performance degradation. Note that Hybrid-DOP still remains suboptimal, as the full potential of hybrid architectures warrants further investigation.

Fig. \ref{fig2}(c) evaluates the proposed algorithms under channel estimation errors defined in (\ref{e24}), where $\sigma _{\text{H}}^{2}\in \left[ 0, 1 \right]$ denotes the CSI uncertainty variance. As $\sigma _{\text{H}}^{2}$ increases from 0 to 1, imperfect CSI hinders MUI suppression, leading to a downward trend of performance similar to the unrefined Hybrid-DOP in Fig. \ref{fig3}(a). Fig. \ref{fig3}(b) shows the spectral efficiency of DOP for each user versus the number of iterations. Since DOP is independent of user ordering, yielding statistically similar performance across users, we use $u=6$ as an example. Simulation results show that the DOP algorithm initialized by EZF exhibits a more rapidly convergence than random initializations. The spectral efficiency of the DOP algorithm increases monotonically and converges rapidly within 10 iterations, as expected from the theoretical analysis in Sec.~\ref{secIV-B}.

\section{Conclusion}
\label{secVI}

This paper proposes an iterative linear DOP algorithm for multiuser beamforming. By alternating between MUI elimination and combiner refinement, the proposed DOP scheme ensures a monotonic increase in spectral efficiency, closely approaching the theoretical optimum while outperforming existing linear benchmarks. The potential of DOP in emerging technologies, such as holographic MIMO communication \cite{gong2023holographic}, remains a promising avenue for future investigation.

\bibliographystyle{IEEEtran}
\bibliography{references}

\vfill

\end{document}